\documentclass[epj]{webofc}
\usepackage[utf8]{inputenc}
\usepackage[varg]{txfonts}   
\usepackage{booktabs}
\usepackage{xcolor}
\definecolor{darkred}{rgb}{0.4,0.0,0.0}
\definecolor{darkgreen}{rgb}{0.0,0.4,0.0}
\definecolor{darkblue}{rgb}{0.0,0.0,0.4}
\usepackage[bookmarks,linktocpage,colorlinks,
    linkcolor = darkred,
    urlcolor  = darkblue,
    citecolor = darkgreen]{hyperref}
\usepackage{amsmath,amssymb}
\usepackage{graphicx}
\usepackage{subfigure}
\usepackage{dsfont}

\wocname{EPJ Web of Conferences}
\woctitle{Lattice2017}
\newcommand{\mpi}{M_{\pi^\pm}}
\newcommand{\mpin}{M_{\pi^0}}

\newcommand{\meta}{M_{\eta_2}}

\newcommand{\fig}[1]{Figure~\ref{#1}}
\newcommand{\tab}[1]{Table~\ref{#1}}
\newcommand{\eq}[1]{Eq.~\ref{#1}}

\begin{document}
%
\selectlanguage{english}
\title{%
The $\eta^\prime$ meson at the physical point with $N_f=2$ Wilson twisted mass fermions
}
\author{%
  \firstname{Christopher} \lastname{Helmes}\inst{1} \and
  \firstname{Bastian} \lastname{Knippschild}\inst{1} \and
  \firstname{Bartosz}  \lastname{Kostrzewa}\inst{1} \and
  \firstname{Liuming} \lastname{Liu}\inst{1} \and
  \firstname{Christian} \lastname{Jost}\inst{1} \and
  \firstname{Konstantin} \lastname{Ottnad}\inst{1}\fnsep\inst{2} \and
  \firstname{Carsten} \lastname{Urbach}\inst{1}\fnsep\thanks{Speaker, \email{urbach@hiskp.uni-bonn.de}} \and
  \firstname{Urs} \lastname{Wenger}\inst{3} \and
  \firstname{Markus} \lastname{Werner}\inst{1}
  \\ for the ETM collaboration
}
\institute{%
  HISKP (Theory) and Bethe Center for Theoretical Physics, University of Bonn, Germany
  \and
  Institut für Kernphysik, Johannes Gutenberg-Universität Mainz, Germany
  \and
  ITP, Albert Einstein Center for Fundamental Physics, University of Bern, Switzerland
}
\abstract{%
  We present results for the $\eta^\prime$ meson and the topological
  susceptibility in $N_f=2$ flavour lattice QCD. The results are obtained
  using Wilson twisted mass fermions at maximal twist with pion masses
  ranging from $340\ \mathrm{MeV}$ down to the physical point. A comparison to
  literature values is performed giving a handle on discretisation
  effects.
}
\maketitle
\section{Introduction}\label{intro}

Due to the persisting $3-5\ \sigma$ deviation in the anomalous
magnetic moment of the muon $a_\mu$ between theory and experiment
there is
considerable interest in the decays $\eta\to\gamma^\star\gamma^\star$
and $\eta^\prime\to\gamma^\star\gamma^\star$. A better knowledge of
the corresponding transition form factors could help to reduce the
uncertainty in the hadronic light-by-light contribution to
$a_\mu$, see for instance Ref.~\cite{Jegerlehner:2015stw}.
Moreover, $\eta$ and $\eta^\prime$ mesons are interesting from a
theoretical point of view because the large mass of the $\eta^\prime$
meson is explained by the anomalously broken $U_A(1)$ axial symmetry
in QCD. The $\eta,\eta^\prime$ mixing pattern and the aforementioned
transition form factors can be computed non-perturbatively using
lattice techniques.
There has been considerable progress in studying $\eta$ and
$\eta^\prime$ mesons from lattice QCD. In Ref.~\cite{Michael:2013gka}
the corresponding mixing has been studied for three values of the
lattice spacing and a large, but still unphysical range of pion mass
values in $N_f=2+1+1$ flavour QCD. After extrapolation to the physical
pion mass value excellent agreement to experiment was found. Further
lattice results for $\eta,\eta^\prime$ can be found in
Refs.~\cite{Christ:2010dd,Dudek:2011tt,Gregory:2011sg,Dudek:2013yja},
however, all at a single lattice spacing value and unphysically large
pion mass values. In Refs.~\cite{Fukaya:2015ara,Aoki:2017paw} the
continuum limit has been studied.

In this proceeding we attempt to study the $\eta^\prime$ meson directly at
the physical point, in a first step in $N_f=2$ flavour QCD. In
$N_f=2$ flavour QCD there exists a pion triplet and one flavour
singlet, which is related to the aforementioned anomaly. We will denote
it as $\eta_2$ meson to distinguish it from the $\eta^\prime$ meson in
full QCD, which is only approximately a flavour eigenstate. The
$\eta_2$ and the $\eta^\prime$ meson have in common that both receive
significant fermionic disconnected contributions to their
correlation functions. Their masses are
expected to differ only by $200\ \mathrm{MeV}$, since the strange
quark is expected to contribute significantly less than the light
quarks~\cite{McNeile:2000hf}. In particular, both are expected to have
a similar dependence on the light quark mass. The most recent lattice
QCD study of the $\eta_2$ meson can be found in
Ref.~\cite{Jansen:2008wv}.

Hence, studying the $\eta_2$ meson at the physical point will reveal
on the one hand important qualitative information on the
realisation of the anomaly in QCD. On the other hand it represents
a feasibility study for a later investigation of $\eta$ and
$\eta^\prime$ in $N_f=2+1+1$ QCD at the physical
point~\cite{finkenrath:2017lat}. The results 
obtained here are also important prerequisites for an exploratory study
of $\eta_2\to\gamma^\star\gamma^\star$.

Since the $\eta_2$ meson is tightly connected to topology, we augment
this study by measuring the topological susceptibility as well. For
this purpose we apply gradient flow techniques~\cite{Luscher:2010iy}
and compare to results available in the literature.

\section{Lattice Action and Operators}\label{sec-1}

\begin{table}[t!]
 \centering
 \caption{The gauge ensembles used in this study. The labelling of
   the ensembles follows the notations in
   Ref.~\cite{Abdel-Rehim:2015pwa}. In addition to the relevant input
   parameters we give the lattice volume $(L/a)^3\times T/a$ and  the
   number of evaluated configurations $N_\mathrm{conf}$.}
 \label{tab:setup}
 \begin{tabular*}{.9\textwidth}{@{\extracolsep{\fill}}lcccccr}
   \toprule
   ensemble & $\beta$ & $c_{\mathrm{sw}}$ &$a\mu_\ell$  &$(L/a)^3\times
   T/a$ & $N_\mathrm{conf}^{\eta_2}$ & $N_\mathrm{conf}^{\chi_t}$  \\
   \midrule
   $cA2.09.48$ &2.10 &1.57551 &0.009  &$48^3\times96$ & $615$ & $1277$\\
   $cA2.30.48$ &2.10 &1.57551 &0.030  &$48^3\times96$ & $352$ & $702$\\
   $cA2.30.24$ &2.10 &1.57551 &0.030  &$24^3\times48$ & $-$   & $703$
   \\                                                                                                                                                                                  
   $cA2.60.32$ &2.10 &1.57551 &0.060  &$32^3\times64$ & $337$ & $2590$\\
   $cA2.60.24$ &2.10 &1.57551 &0.060  &$24^3\times48$ & $-$   & $1611$\\
  \bottomrule
 \end{tabular*}

\end{table}

The results presented in this paper are based on the gauge
configurations generated by the ETMC with Wilson clover twisted mass
quark action at maximal
twist~\cite{Frezzotti:2000nk}. We employ the Iwasaki gauge
action~\cite{Iwasaki:1985we}.
The measurements are performed on up to five $N_f = 2$ ensembles with
pion mass at its physical value, 240~MeV and 340~MeV,
respectively. The lattice spacing is $a=0.0931(2)\ \mathrm{fm}$ for
all three ensembles.  In Table~\ref{tab:setup} we list the five
ensembles with the relevant input parameters, the lattice volume and
the number of configurations. More details about the ensembles can be
found in Ref.~\cite{Abdel-Rehim:2015pwa}.

As a smearing scheme we use the stochastic Laplacian Heavyside (sLapH)
method~\cite{Peardon:2009gh,Morningstar:2011ka} for our
computation. The details of the sLapH parameter choices for a set of
$N_f = 2+1+1$ Wilson twisted mass ensembles are given in
Ref.~\cite{Helmes:2015gla}. The parameters for the ensembles used in
this work are the same as those for $N_f = 2+1+1$ ensembles with the
corresponding lattice volume.

In $N_f=2$ flavour QCD there is the neutral pion, corresponding to the
neutral one of the three pions in the triplet, and the $\eta_2$, the
flavour singlet pseudo-scalar meson related to the axial $U(1)$
anomaly. The (hermitian) interpolating operator projected to zero
momentum reads
\begin{equation}
  \mathcal{O}^s(t)\ =\ \frac{1}{\sqrt{2}}\sum_\mathbf{x} \bar\psi(\mathbf{x},t)
  i\gamma_5\,\mathds{1}_f\, \psi(\mathbf{x},t)\,.
\end{equation}
Here, $\mathds{1}_f$ is the unit matrix acting in flavour space. From
the operator one builds the correlation functions
\begin{equation}
  \label{eq:Cor}
  C_{\eta_2}(t-t')\ =\ \langle \mathcal{O}^s(t)\,
  (\mathcal{O}^s(t'))^\dagger \rangle\,,
\end{equation}
which allows one to determine the mass $\meta$ from its
decay in Euclidean time. 

The correlation function in \eq{eq:Cor} has both a fermionic connected
and a fermionic disconnected contribution. 
The connected contribution can be estimated with high statistical
accuracy. It decays asymptotically exponentially with the mass of the
(connected neutral) pion. The disconnected part is statistically
noisy. The signal for the $\eta_2$ meson arises from the difference of
connected and disconnected contribution, 
\begin{equation}
  \label{eq:trcorr}
  C_{\eta_2}(t)\ =\ C^\mathrm{conn}(t)\ -\ C^\mathrm{disc}\,,
\end{equation}
where the disconnected contribution has to cancel the pion contribution coming from
the connected part. Since the pion is significantly lighter than the
$\eta_2$, the statistical analysis is rather delicate. 

The topological susceptibility is a measure for the fluctuations of the
topological charge. It is defined as
\begin{equation}
  \chi_t\ =\ \frac{1}{V} \int \mathrm{d} x\ \int \mathrm{d} y\ \langle
  q(x) q(y)\rangle 
\end{equation}
with the volume $V$ and the topological charge density $q(x)$ for
which we use a clover-type discretisation of the field strength 
tensor, following the conventions in Ref.~\cite{Bruno:2014ova}.  For
the topological susceptibility $\chi_t$ we employ the gradient flow as
introduced for lattice QCD in Ref.~\cite{Luscher:2010iy}. This setup
has the advantage of yielding a renormalised topological
susceptibility at any finite flow time $t$. Since the renormalised
susceptibility is scale invariant, it becomes independent of $t$ at
sufficiently large flow times. This is indeed what we observe in our
calculation and we choose $t=3 t_0$ with $\sqrt{t_0}=0.1535(12)$\,fm
from Ref.~\cite{Abdel-Rehim:2015pwa} obtained at the physical point.

\section{Analysis}

The data is analysed using a blocked bootstrap procedure with $R=1500$
bootstrap samples. Depending on the ensemble, we have chosen the block
size such that roughly $150$ blocked data points where left. For the
analysis of the topological susceptibility we use $R=1000$ bootstrap
samples and block sizes yielding around $100$ data blocks depending on
the ensemble under consideration. The resulting error is compared to
the naive one appropriately rescaled with the integrated
autocorrelation time of $Q^2$, and the larger of the two is chosen as
the final error.

\subsection{Excited State Subtraction}

In particular for the $\eta_2$ meson, the fermionic disconnected
contributions are very noisy. As a consequence, the signal is lost
relatively early in Euclidean time. For this reason we have in the
past applied a method to subtract excited
states~\cite{Jansen:2008wv,Michael:2013gka,Liu:2016cba}, originally
proposed
in Ref.~~\cite{Neff:2001zr}. Because it works very well, we will
apply it here again for the $\eta_2$ meson. It consists of subtracting
excited states from the connected contribution only. This is feasible,
because the connected part -- representing a pion correlation
function -- has a signal for all Euclidean time values. Therefore, we
can fit it at large enough Euclidean times such that excited states
have decayed sufficiently. Next, we replace the connected correlation
function at small times by the fitted function. Thereafter, the such
subtracted connected contribution is summed according to
\eq{eq:trcorr} to the full $\eta_2$ correlation functions.

The underlying assumption is that disconnected contributions are large
for the ground state, i.e. the $\eta_2$, but not for excited
states. If this assumption is correct, the effective mass
\begin{equation}
  M_\mathrm{eff}\ =\ -\log\frac{C_{\eta_2}(t)}{C_{\eta_2}(t+1)}
\end{equation}
should show a plateau from very early Euclidean times on. We have
found in Refs.~\cite{Jansen:2008wv,Michael:2013gka} that this approach
works very well for the $\eta_2$ meson in $N_f=2$ flavour QCD as well
as for $\eta$ and $\eta^\prime$ mesons in $N_f=2+1+1$ flavour QCD.

Phenomenologically, the connected contribution gives the pion as the
ground state for large Euclidean times. And, due to symmetries, it
cannot couple to the $\eta_2$. The pion contribution needs to be
cancelled by the disconnected contribution, which, therefore, reads
\[
C_{\eta_2}^\mathrm{disc}\ \approx\ A\cdot e^{-M_{\pi^0}}
- B\cdot e^{-M_{\eta_2} t}
\]
if no higher excited states contribute. 

\subsection{Shifted Correlation Functions}

The expected time dependence of the correlation functions considered
here reads as follows
\begin{equation}
  C(t) = |\langle 0|\mathcal{O}^s|0\rangle|^2 + \sum_n \frac{|\langle
    0| \mathcal{O}^s| n\rangle|^2}{2E_n}\left(e^{-E_n t} + e^{-E_n
    (T-t)}\right)\,, 
\end{equation}
where $n$ labels the
states with the corresponding quantum numbers. The first, time
independent term on the right hand side corresponds to the vacuum
expectation value (vev). Using the symmetries of our action one can
show that for $\mathcal{O}^s$ the vev must be zero. Potential
deviations from this expectation due to finite statistics and volume
can be dealt with by building the shifted correlation function
\begin{equation}
  \label{eq:corrshifted}
  \tilde{C}(t)\ =\ C(t) - C(t+1)\,.
\end{equation}
The difference cancels any constant contributions in the correlation
function, while also
changing the time dependence to be anti-symmetric in time
\begin{equation}
  \tilde{C}(t)\ \propto\ \left(e^{-E_n t} - e^{-E_n (T-t)}\right)\,.
\end{equation}
In addition, there can be finite volume effects due to incorrectly
sampled topological charge sectors. The corresponding formulae for
fixed topological sectors have been worked out in
Ref.~\cite{Aoki:2007ka}. They lead to a finite volume effect in 
$C_{\eta_2}$ constant in Euclidean time of the form~\cite{Bali:2014pva}
\begin{equation}
  \propto\ \frac{a^5}{T}\left(\chi_t - \frac{Q^2}{V}\right)
\end{equation}
proportional to the difference of the topological susceptibility,
$\chi_t$, and the expectation value of the squared topological
charge $Q^2/V$. 
If present, such a term will cause the $\eta_2$ correlation function to
stay finite at large Euclidean times. Depending on the sign of the
coefficient in front of the finite volume effect, the correlation
function may even turn negative at relatively small Euclidean times.
Clearly, a finite volume effect of this type can be subtracted again
using the shifting procedure.

The shifting procedure has the additional advantage of removing
statistical correlations between neighbouring time slices $t/a$ and
$t/a+1$. Therefore, shifting can also be advantageous for the analysis
procedure, because the inverse variance-covariance matrix is often
difficult to estimate reliably~\cite{Michael:1994sz}. Of course, when
the correlation between adjacent time slices can be taken into account
correctly, there is no difference in using the unshifted or the
shifted correlation function and both give exactly identical results
(unless there is a constant contribution).

\subsection{Scale Setting}

In order to compare the $\eta_2$ masses to older publications
available in the literature, we use the Sommer scale $r_0/a$ for the
$\meta$. The value $r_0/a=5.317(48)$ from
Ref.~\cite{Abdel-Rehim:2015pwa} for the $cA2.09.48$ ensemble at the
physical point is used for all ensembles considered here. The corresponding
physical value reads $r_0=0.4907(5)\ \mathrm{fm}$.

Since we use the gradient flow to determine the topological
susceptibility, it is natural to use a scale from the gradient flow
for $\chi_t$. In order to compare our results with the ones from
Ref.~\cite{Bruno:2014ova} we use the scale $t_1(m_\pi)$. Its physical
value at the physical point reads $t_1 = 0.061\
\mathrm{fm}^2$~\cite{Bruno:2014ova}. Note, however, that $\chi_t$ is evaluated
at flow time $t=3 t_0$ as discussed above.

\section{Results}

\begin{figure}[thb]
  \centering
  \subfigure{\includegraphics[width=.45\linewidth,page=3]{cA2_30_48all}}
  \subfigure{\includegraphics[width=.45\linewidth,page=1]{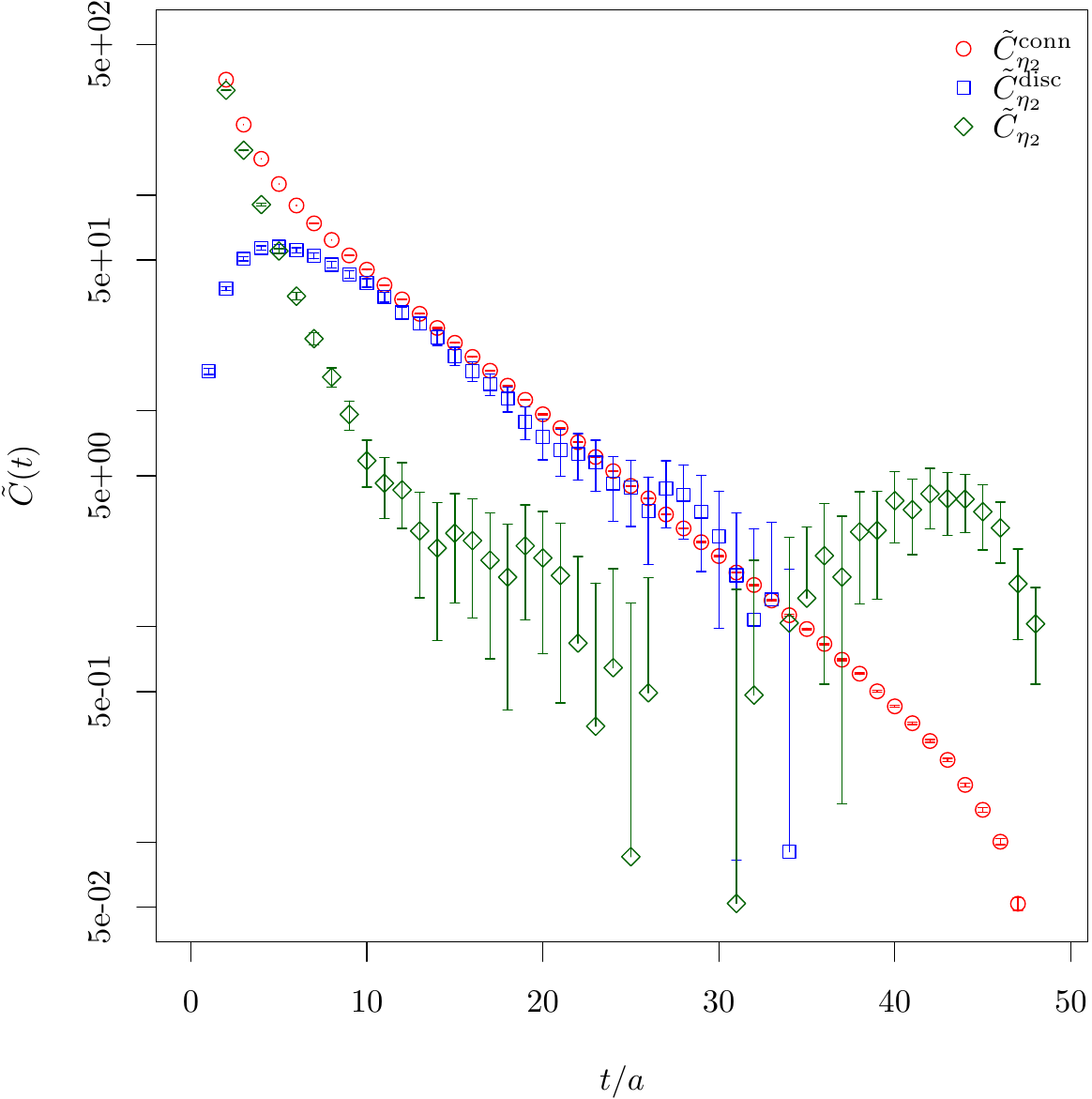}}
  \caption{Left: $\eta_2$ correlation function $C_{\eta_2}$ and its
    shifted counterpart $\tilde{C}_{\eta_2}$ for ensemble cA2.30.48 as
  a function of $t/a$. Right: connected, disconnected and full shifted
  correlation function $\tilde{C}_{\eta_2}$ as a function of $t/a$
  with logarithmic $y$-axis.}
  \label{fig:corr}
\end{figure}

In the left panel of \fig{fig:corr} we show the $\eta_2$ correlation
function $C_{\eta_2}$ and the corresponding shifted one
$\tilde{C}_{\eta_2}$ as functions of the Euclidean time for ensemble
cA2.30.48. For $C_{\eta_2}$ errors are larger than the typical
fluctuations for $t/a>10$, pointing towards large correlation between
separate time slices. Still, the correlation function is not
incompatible with zero at large times. In the shifted correlation
function $\tilde{C}_{\eta_2}$ the correlation is largely removed and
the error bars are of the typical fluctuation size. Also
$\tilde{C}_{\eta_2}$ is compatible with zero for large times
$t/a>15$. 

In the right panel of \fig{fig:corr} we show the different
contributions to the $\eta_2$ shifted correlation function for
ensemble cA2.30.48 in a logarithmic plot. While the signal for the
connected part (red circles) is excellent up to $t/a=48$, the
disconnected contribution (blue squares) starts to become noisy around
$t/a=15$. We observe that connected and disconnected contributions are
very similar in magnitude from around $t/a=10$ on. The full $\eta_2$
correlator (green diamonds) is given by the difference of connected
and disconnected parts. Consequently, the signal deteriorates around
$t/a=10$.

From the left panel of \fig{fig:corr} it is also clear that the ground
state energy and amplitude of the connected contribution (in twisted
mass the connected neutral pion) can be extracted reliably. The connected
contribution is then replaced by only this ground state and the full
disconnected part is subtracted from it. As a result we obtain
plateaus from $t/a=3$ on. 

\begin{table}[t!]
 \centering
 \caption{Results for the $\eta_2$ meson in lattice units and for the
   topological susceptibility in units of $t_1^2$ for the ensembles
   investigated here.}
 \begin{tabular*}{.6\textwidth}{@{\extracolsep{\fill}}lcc}
  \toprule
  Ensemble   & $a\meta$ & $10^3 t_1^2\ \chi_t$ \\
  \midrule
  $cA2.09.48$  &  $0.358(9)$ & $0.53(4)$\\
  $cA2.30.48$  &  $0.372(8)$ & $0.63(6)$\\
  $cA2.30.24$  &  ---        & $0.52(6)$\\
  $cA2.60.32$  &  $0.374(7)$ & $0.91(5)$\\
  $cA2.60.24$  &  ---        & $0.91(8)$\\
  \bottomrule
 \end{tabular*}
 \label{tab:results}
\end{table}

\begin{figure}[thb] 
  \centering
  \includegraphics[width=.8\linewidth,page=1,clip]{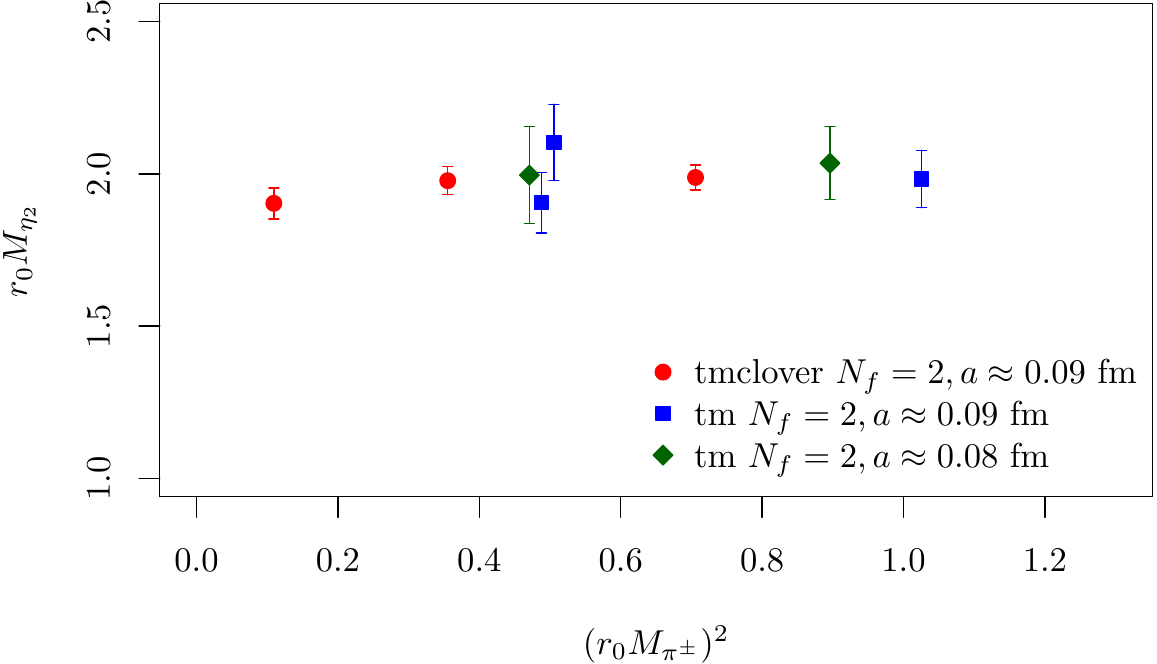}
  \caption{Compilation of $N_f=2$ Wilson twisted mass results with and
    without clover term for the $\eta_2$ meson. The twisted mass (tm)
    results have been taken from Ref.~\protect{\cite{Jansen:2008wv}}.}
  \label{fig:r0eta2}
\end{figure}

The such extracted energy levels in lattice units are compiled in
\tab{tab:results}. The $\eta_2$ masses are shown in units of the
Sommer parameter $r_0$ in \fig{fig:r0eta2} as a function of the
squared pion mass $(r_0 \mpi)^2$. The red circles represent the
results presented here. In addition we show results
for Wilson twisted mass fermions with $N_f=2$ dynamical quark flavours
without clover term taken from Ref.~\cite{Jansen:2008wv}. The blue
squares correspond to a lattice spacing of about $0.09\ \mathrm{fm}$,
while the green diamonds correspond to about
$0.08\ \mathrm{fm}$. These results from Ref.~\cite{Jansen:2008wv} have
larger statistical errors and the smallest pion mass corresponds to
about $300\ \mathrm{MeV}$. However, the agreement with the results
presented here is good. In particular, the almost constant
extrapolation to the physical point is confirmed. 
An estimate of lattice artefacts is difficult, but they seem to be not
larger than our statistical uncertainties.

\begin{figure}[thb] 
  \centering
  \sidecaption
  \includegraphics[width=.6\linewidth,page=2,clip]{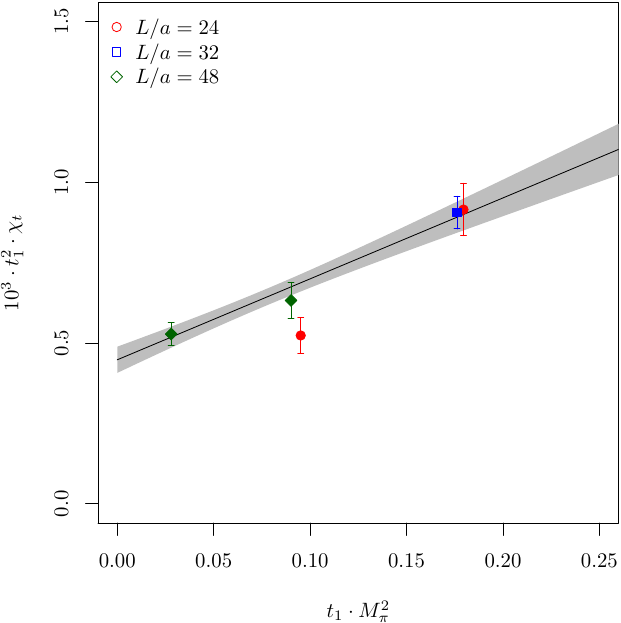}
  \caption{Topological susceptibility $\chi_t$ as a function of the
    squared pion mass, both in appropriate units of $t_1$. The solid
    line with shaded error band indicates a fit to the data
    according to \eq{eq:topsus}.}
  \label{fig:chit}
\end{figure}

In \tab{tab:results} we have also compiled our results for the
topological susceptibility in units of $t_1$. They are displayed in
\fig{fig:chit} as a function of $t_1 \mpi^2$. 
In leading order Wilson ChPT one expects the following dependence of
$\chi_t$ on the lattice spacing and the pion mass~\cite{Bruno:2014ova}
\begin{equation}
  \label{eq:topsus}
  t_1^2\, \chi_t\ =\ \frac{1}{8}\, t_1^2\, f_\pi^2\, \mpin^2\ +\ a^2\, \frac{c_2}{t_1}\,.
\end{equation}
Apart from the ensemble with too small volume cA2.30.24, our data is
nicely compatible with this expectation: the solid line in
\fig{fig:chit} represents a fit of the function
\[
g(\mpi^2)\ =\ c_1 t_1 \mpi^2\ +\ a^2\, \frac{c_2}{t_1}
\]
to the data with fit parameters $c_1$ and $c_2$. Note that we use the
charged pion mass, because charged and neutral pion masses are
degenerate within errors~\cite{Abdel-Rehim:2015pwa}. The best fit
parameter for $c_1$ is compatible with $t_1f_\pi^2/8$. Note that
ensemble cA2.30.24 has a very small volume explaining the outlier in
\fig{fig:chit}. 

The fitted value for $c_2$ can be compared to the results of
Ref.~\cite{Bruno:2014ova}. We obtain $c_2=3.3(3)\cdot10^{-3}$, while
Bruno et al. have $c_2 = 5.1(7)\cdot 10^{-3}$. 

\section{Summary}

In this proceeding contribution we have presented results for the
$\eta_2$ meson related to the axial anomaly and the topological
susceptibility in two-flavour QCD. The 
results have been obtained using $N_f=2$ lattice QCD ensembles
generated by ETMC with the Wilson twisted clover
discretisation~\cite{Abdel-Rehim:2015pwa}. Pion mass values reach from
its physical value up to $340\ \mathrm{MeV}$ at a single lattice
spacing value of $a=0.0931(2)\ \mathrm{fm}$. For the $\eta_2$ we could
confirm the almost constant extrapolation in $\mpi^2$ towards the
physical point. Errors are significantly reduced compared to previous
calculations. Lattice artefacts seem to be not larger than our
statistical uncertainty. 

The topological susceptibility has been computed using the gradient
flow. As expected, $\chi_t$ is proportional to $\mpi^2$ up to a
lattice artefact independent of $\mpi$. Even if we are not able to
finally confirm this with a single lattice spacing at hand, this constant term
should be of $\mathcal{O}(a^2)$. The size of this artefact appears to
be significantly smaller than what is observed with Wilson clover
fermions in Ref.~\cite{Bruno:2014ova}. 

We thank the members of ETMC for the most enjoyable
collaboration. This project was funded by the DFG as a project in the
Sino-German CRC110. The computer time for this project was made
available to us by the John von Neumann-Institute for Computing (NIC)
on the Juqueen system in J{\"u}lich and by the ITP Bern on their HPC
clusters.
This work was granted access to the HPC resources of CINES and IDRIS
under the allocation 52271 made by GENCI.

\bibliography{bibliography}

\end{document}